\documentclass[notitlepage,a4paper,aps,prd,onecolumn,superscriptaddress,nofootinbib,groupedaddress]{revtex4-2}

\usepackage{amsmath}
\usepackage{amsfonts,color}
\usepackage{amsmath}
\usepackage{amsthm}
\usepackage{amsmath}
\usepackage{empheq}
\usepackage{amsfonts,color}
\usepackage{amssymb,float}
\usepackage{booktabs,multirow}
\usepackage{titlesec}
\usepackage{braket}

\usepackage[ignoreall]{geometry}

\usepackage[en-US]{datetime2}
\DTMsettimestyle{iso}
\DTMsetup{showseconds=false}

\titleformat{\paragraph}
{\normalfont\normalsize\bfseries}{\theparagraph}{1em}{}
\titlespacing*{\paragraph}
{0pt}{3.25ex plus 1ex minus .2ex}{1.5ex plus .2ex}

\usepackage{amsmath}
\usepackage[utf8]{inputenc}
\allowdisplaybreaks
\usepackage[T1]{fontenc}
\usepackage{color}
\usepackage{hyperref}
\hypersetup{
    colorlinks=true,
    linkcolor=blue,
    filecolor=magenta,      
   citecolor=blue
}

\usepackage{enumitem,accents}

\usepackage{tikz}
\usetikzlibrary{shapes.geometric}
\usetikzlibrary{arrows.meta,arrows}

\usepackage[normalem]{ulem}

\newcommand{\lc}[1]{\accentset{\circ}{#1}}

\newcommand{\ud}[3]{#1^{#2}_{\phantom{#2}#3}}
\newcommand{\du}[3]{#1^{\phantom{#2}#3}_{#2}}

\definecolor{vert}{rgb}{0.,0.65,0.}

\newcommand{\needref}[1][]{ \text{{\color{red} [/!$\backslash$ Add reference\ifstrequal{#1}{}{ }{ #1 }/!$\backslash$]}}}
\newcommand{\needdetails}[1][]{ \text{{\color{red} [/!$\backslash$ Add details\ifstrequal{#1}{}{ }{ #1 }/!$\backslash$]}}}

\RequirePackage{mathrsfs}
\newcommand{\lagrangian}{\mathcal{L}}

\newcommand{\dint}{\mathrm{d}}
\newcommand{\diff}{\mathrm{d}}
\newcommand{\man}[1][M]{#1}
\newcommand{\Lie}{\mathfrak{L}}

\newcommand{\tens}[1]{\mathbf{#1}}
\newcommand{\tensi}[1]{\boldsymbol{#1}}
\newcommand{\metric}{\boldsymbol{g}}
\newcommand{\curve}[1][C]{\mathscr{#1}}

\DeclareMathOperator{\diag}{diag}

\DeclareMathOperator{\met}{Met}
\DeclareMathOperator{\tor}{Tor}

\begin{document}
\title{Metric-affine cosmological models and the inverse problem of the
  calculus of variations.\\ Part II: Variational bootstrapping of the $\Lambda$CDM model}

\author{Ludovic Ducobu}
\email{ludovic.ducobu@umons.ac.be}
\affiliation{Department of Mathematics and Computer Science, Transilvania University of Brasov, Brasov, Romania}
\affiliation{Nuclear and Subnuclear Physics, University of Mons, Mons, Belgium}

\author{Nicoleta Voicu}
\email{nico.voicu@unitbv.ro}
\affiliation{Department of Mathematics and Computer Science, Transilvania University of Brasov, Brasov, Romania}
\affiliation{Lepage Research Institute, Presov, Slovakia}


\begin{abstract}
The method of variational bootstrapping, based on canonical variational completion, allows one to construct a Lagrangian for a physical theory depending on two sets of field variables, starting from a guess of the field equations for only one such set. This setup is particularly appealing for the construction of modified theories of gravity, since one can take lessons from GR for an “educated guess” of the metric field equations; the field equations for the other fields are then fixed by the obtained Lagrangian (up to terms that are completely independent from the metric tensor).

In the present paper, we apply variational bootstrapping to determine metric-affine models which are, in a variational sense, closest to the $\Lambda$CDM model of cosmology. Starting from an ``educated guess'' that formally resembles the Einstein field equations with a cosmological ``constant'' (actually, a scalar function built from the metric and the connection) and a dark matter term, the method then allows to find ``corrected'' metric equations and to ``bootstrap'' the connection field equations.

Lagrangians obtained via this method, though imposing some restricting criteria, still encompass a wide variety of metric-affine models. In particular, they allow for a subclass of quadratic metric-affine theories restricted to linear terms in the curvature tensor.

\end{abstract}

\maketitle


\section{Introduction}

Despite its successes, it is notorious that General Relativity (GR) struggles at the smallest scales -- due to an incompatibility with quantum theory -- and at the largest scales -- where it fails to correctly reproduce the cosmological behavior of the universe. For the latter problem (which is the one we focus on, in the present paper), a simplest way of correcting the GR model is to amend Einstein's equation
\begin{equation}
\label{eq:Einstein}
G_{\mu\nu} = \frac{8\pi\mathcal{G}}{c^4} T_{\mu\nu}
\end{equation}
by adding entities known as dark matter and dark energy. This paradigm is known as the $\Lambda$CDM model of cosmology \cite{Riess_1998, Perlmutter_1999, Planck:2018vyg, weinberg1972gravitation, 10.1093/lambdaCDMtextbook} and is, to this date, the best model available to fit cosmological observations. Nevertheless, the $\Lambda$CDM model is challenged, on the one hand, by the so-called $H_0-\sigma_8$ tension \cite{Bernal_2016} and on the other hand, by the fact that the very physical origin of dark energy and dark matter physical origin still eludes our understanding. These puzzles legitimize the quest for a more general (classical) gravitational theory.

Such generalizations of GR concern the geometric structure of spacetime (i.e., the \textit{kinematical} level), the field equations (in other words, the \textit{dynamical} level), or both.

\vspace{10pt}
At the kinematical level, one of the simplest extensions is metric-affine geometry, which allows for more freedom by relaxing the GR hypotheses on the spacetime connection (metricity, absence of torsion), \cite{Blagojevic:2012bc, HEHL19951, CANTATA:2021ktz}. In metric-affine gravity models, one can thus encode the dark sector into the ``enhanced'' geometry of spacetime\footnote{Working directly at the dynamical level -- without necessarily generalizing the spacetime structure -- can be done by adding new fields in the theory such as, \emph{e.g.}, in scalar-tensor gravity \cite{Horndeski1974}.}.

\vspace{10pt}
At the dynamical level, one usually discusses modified theories of gravity by first postulating a Lagrangian, from which the field equations are then derived. Even though this procedure is automatically justified as long as one expects a fundamental physical theory to admit a Lagrangian, ultimately what one typically wants to control are the properties of the \textit{field equations}, rather than the ones of the Lagrangian. For this reason, one of us proposed in \cite{Voicu_2015}, together with D. Krupka, a systematic method, called \textit{canonical variational completion}, allowing to go the other way around: starting with an approximate form (an ``educated guess'') of the field equations, one can canonically construct a Lagrangian and subsequently, the ``corrected'' field equations. The obtained field equations are the closest ones to our original guess, in the sense that the correction terms to be added specifically measure the ``obstruction from variationality'' of this guess.  
The method of canonical variational completion was then upgraded in the first part of this work \cite{Ducobu:2024grm}, to accommodate theories that involve more than one set of dynamical variables, e.g., a metric and a connection. We showed that from an ``educated guess'' of the field equations with respect to one variable (e.g., the metric) only, one can still canonically ``bootstrap'' a Lagrangian, up to boundary terms and to terms that are completely independent of the respective variable.   

\vspace{10pt}
In the present paper, we apply the method proposed in \cite{Ducobu:2024grm} to select those metric-affine Lagrangians producing the ``closest to $\Lambda$CDM'' metric equations. 
Our ``educated guess'' (or ``seed'') for the metric equations is a system that formally resembles the Einstein equations with a cosmological constant (encoding dark energy, $w=-1$) and a term describing dark matter (with $w=0$), constructed following several ideas which we list below:
\begin{enumerate}
\item \textit{General covariance.} The seed equations must be generally covariant.
\item \textit{The metric-affine hypothesis.} The dark sector has a geometric (more precisely, metric-affine) origin.\\
The above, together with general covariance, imply that the cosmological ``constant'' is actually a scalar function depending on the metric and on the connection components; similarly, the dark matter term is a tensor field which can be completely expressed in terms of the metric and of the connection.
\item \textit{Role of the metric.} The volume form on spacetime and raising/lowering indices of tensors are the standard ones in terms of the spacetime metric tensor.
\item \textit{Minimality of the extension.} By minimality, we will actually mean three different aspects:
 \begin{itemize}
 	\item \textit{Minimal order}: In the seed equations, the dark sector terms are of order at most one.\\
	This ensures that the variationally completed equations are of order at most two, both in the metric and the connection.
	\item \textit{No ad-hoc terms}: To ensure this, we first consider the particular situation of cosmological symmetry -- and in this situation, we determine the most general form of (0,2)-type tensor fields fitting the dark energy, respectively, dark matter equations of state (EOS)\footnote{Actually, we point out a much stronger statement: any cosmologically symmetric (0,2)-type tensor field can be split into a sum of two terms which formally obey the equations of state of dark energy, respectively, of dark matter.}. Next, stepping back from cosmological symmetry, we consider the sum of the two obtained tensor fields (without adding any other, \textit{``ad hoc''}, terms) as our ``educated guess'', for the fully general case.
	\item \textit{The finite sum ansatz}: Each of the dark sector terms can be expressed as a finite sum of homogeneous expressions (not necessarily of positive degrees) in the metric and its partial derivatives.\\
	To make sense of the last hypothesis, we must bear in mind that the metric is used in the dark sector terms in three ways: to construct connection coefficients/curvature components, to raise indices, respectively, to construct the volume form. Each of these operations employs a finite number of metric tensor components or partial derivatives thereof.   
	
\end{itemize}  
\end{enumerate} 

The above hypotheses ensure, on the one hand, that the obtained Lagrangians are generally covariant and, on the other hand, that \textit{all} their non-boundary terms can be found from our ``educated guess'' (or seed) of the metric equations only. This way, no input on the connection equations is necessary; these can be automatically ``bootstrapped'' from the obtained Lagrangian. 

We find that a number of the existing metric-affine models do satisfy the above minimality requests, \emph{e.g.}, those whose Lagrangian is the Ricci scalar of the independent connection and a subclass of quadratic metric-affine theories avoiding non-linear terms in the curvature tensor of the  connection.

\vspace{10pt}
The paper is structured as follows: In sections \ref{sec:setup} and \ref{sec:LCDMandco}, we fix our notations for the metric-affine geometry and briefly review some properties of the $\Lambda$CDM model that are necessary for our discussion. In section \ref{sec:extendLCDM}, we consider metric-affine theories of gravity based on modifications implemented directly at the level of GR field equations, with the aim of obtaining an ``educated guess'' for the metric field equation in a metric-affine extension of GR. Section \ref{sec:bootstrap_lambdaCDM} contains our main results, as follows. In subsection \ref{sec:userguideVB}, we remind the basics of the technique of variational bootstrapping presented in \cite{Ducobu:2024grm}. Then, in subsection \ref{sec:hypothesis}, we explain our hypotheses in the construction of our ``educated guess'', before performing, in subsection \ref{sec:VBLCDM}, its variational bootstrapping, leading to a Lagrangian (actually, a class of Lagrangians) for the full theory. Then, in subsection \ref{sec:fieldeqVBLCDM}, we determine the field equations for the obtained Lagrangians. In subsection \ref{sec:Analysiseq}, we then comment on the obtained equations and on their link with previously formulated metric-affine models of gravity. Finally, in section \ref{sec:ccl}, we present our conclusions and discuss further directions of research.

\section{Geometric setup}\label{sec:setup}

In the following, we model spacetime by means of a triple $\left(\man, \Gamma, \metric\right)$, where $\man$ is an arbitrary 4-dimensional differentiable manifold, $\Gamma$ is a generic linear connection on $\man$ and $\metric$ a metric with Lorentzian signature. Local coordinates on $\man$ will be denoted by $x^{\mu}$ and the natural (coordinate) local basis vectors on $TM$, by $\partial_\mu$. When writing connection  coefficients, the second lower index refers to the ``derivative index'' meaning that, for example, given a vector field $V = V^\mu \partial_\mu$ one has
\begin{equation*}
\nabla_\mu V^\nu \coloneqq \left(\nabla_{\partial_\mu} V\right)^\nu = \partial_\mu V^\nu + \Gamma^\nu_{\rho\mu} V^\rho.
\end{equation*}

We will denote by $\tens{R}$ the curvature tensor of $\Gamma$:
\begin{equation}
\tens{R}\left(V, W\right)Z \coloneqq \nabla_{V}\nabla_{W}Z - \nabla_{W}\nabla_{V}Z - \nabla_{\left[V, W\right]}Z,
\end{equation}
where $\nabla$ is the covariant derivative associated to $\Gamma$ and by:
\begin{equation}
\ud{R}{\rho}{\sigma\mu\nu} = \partial_\mu\Gamma_{\sigma\nu}^\rho - \partial_\nu\Gamma_{\sigma\mu}^\rho + \Gamma_{\alpha\mu}^\rho\Gamma_{\sigma\nu}^\alpha - \Gamma_{\beta\nu}^\rho\Gamma_{\sigma\mu}^\beta,
\end{equation}
its components in a coordinate basis.
Similarly, $\tens{T}$ will designate the torsion of $\Gamma$:
\begin{equation}
\tens{T}\left(V, W\right) \coloneqq \nabla_{V}W -\nabla_{W}V - \left[V,W\right],
\end{equation}
with components in a coordinate basis:
\begin{equation}
\ud{T}{\rho}{\mu\nu} = \Gamma_{\nu\mu}^\rho - \Gamma_{\mu\nu}^\rho = - 2\ \Gamma_{[\mu\nu]}^\rho,
\end{equation}
and by $\tens{Q}$ the non-metricity tensor defined as
\begin{equation}
\tens{Q} \coloneqq \nabla\metric,
\end{equation}
with local components:
\begin{equation}
Q_{\mu\nu\rho} = \nabla_\rho g_{\mu\nu} \coloneqq \partial_\rho g_{\mu\nu} - \Gamma^\alpha_{\mu\rho}\ g_{\alpha\nu} - \Gamma^\beta_{\nu\rho}\ g_{\mu\beta}.
\end{equation}

\bigskip

Our metric-affine setup can equivalently (and more conveniently, for our purposes) be defined as the triple $\left(\man, \tens{L}, \metric\right)$, where the \textit{distortion} of $\Gamma$:
\begin{equation}
\tens{L}=\Gamma - \lc{\Gamma}
\end{equation}
is a tensor field of type $(1,2)$, with components $\ud{L}{\rho}{\mu\nu}$ in a coordinate basis. The symbol $\circ$ designates the Levi-Civita connection associated to $\metric$ and all its related quantities; for instance:
\begin{equation}
\lc{\Gamma}_{\mu\nu}^\rho \coloneqq \frac{1}{2} g^{\rho\alpha}\left( \partial_\nu g_{\mu \alpha} +\partial_\mu g_{\alpha \nu} - \partial_\alpha g_{\mu\nu} \right), \,\, \Gamma_{\mu\nu}^\rho = \lc{\Gamma}_{\mu\nu}^\rho + \ud{L}{\rho}{\mu\nu}. 
\end{equation}

In the following, for conciseness, we will express our relations in terms of the distortion tensor $\tens{L}$. If needed, all the relations can be translated in terms of the torsion $\tens{T}$ (respectively, the contortion $\tens{K}$) and non-metricity $\tens{Q}$ (respectively, disformation $\tens{D}$) using the relations below, \cite{CANTATA:2021ktz}:
\begin{equation}\label{eq:LfromKandD}
\ud{L}{\rho}{\mu\nu} \coloneqq \ud{K}{\rho}{\mu\nu} + \ud{D}{\rho}{\mu\nu},
\end{equation}
where:
\begin{equation}\label{eq:defD}
\ud{D}{\rho}{\mu\nu} \coloneqq - \frac{1}{2} g^{\rho\alpha} \left(Q_{\nu\alpha\mu} - Q_{\mu\nu\alpha} + Q_{\alpha\mu\nu}\right),
\end{equation}
\begin{equation}\label{eq:defK}
\ud{K}{\rho}{\mu\nu} \coloneqq - \frac{1}{2} \left(\ud{T}{\rho}{\mu\nu} - g_{\nu\alpha}\ g^{\rho\beta}\ \ud{T}{\alpha}{\beta\mu} + g_{\mu\alpha}\ g^{\rho\beta}\ \ud{T}{\alpha}{\nu\beta}\right).
\end{equation}
For the sake of completeness, we also present the inverse relations:
\begin{equation}
\ud{K}{\rho}{\mu\nu} = \ud{L}{\rho}{[\mu\nu]} + g^{\rho\beta}\ \left(g_{\mu\alpha}\ \ud{L}{\alpha}{[\nu\beta]} + g_{\nu\alpha}\ \ud{L}{\alpha}{[\mu\beta]}\right),
\end{equation}
\begin{equation}
\ud{D}{\rho}{\mu\nu} = \ud{L}{\rho}{(\mu\nu)} - g^{\rho\beta}\ \left(g_{\mu\alpha}\ \ud{L}{\alpha}{[\nu\beta]} + g_{\nu\alpha}\ \ud{L}{\alpha}{[\mu\beta]}\right),
\end{equation}
\begin{equation}\label{eq:TfromL}
\ud{T}{\rho}{\mu\nu} = -2\ \ud{L}{\rho}{[\mu\nu]}  = -2\ \ud{K}{\rho}{[\mu\nu]},
\end{equation}
\begin{equation}\label{eq:QfromL}
Q_{\rho\mu\nu} = -2\ g_{(\rho\vert \alpha}\ \ud{L}{\alpha}{\vert\mu)\nu}  = -2\ g_{(\rho\vert \alpha}\ \ud{D}{\alpha}{\vert\mu)\nu},
\end{equation}
where, as usual, square brackets (resp. round brackets) placed around a group of indices denote total antisymmetrization (resp. symmetrization) over those indices, with exception of indices placed between vertical bars.

\section{Brief review of the $\Lambda$CDM model}\label{sec:LCDMandco}

This section briefly presents the restrictions imposed by cosmological symmetry upon tensors on the spacetime manifold and the geometric setup of the $\Lambda$CDM model; for more detail see \cite{weinberg1972gravitation, 10.1093/lambdaCDMtextbook}. Even though most of the material of this section is not new, this allows us to clearly fix our notations and to recall an important -- yet rarely emphasized -- problem.

\subsection{Cosmologically symmetric tensor fields}\label{sec:cosmosymfield}

Cosmological symmetry, understood as invariance under the six Killing vector fields defining it, imposes quite strong constraints on  tensor fields of any rank, which we briefly present below; these six symmetry generators are explicitly displayed in Appendix \ref{sec:cosmosymvec}.
In the following, we denote by $ \left(x^\mu\right) = \left(t, r, \theta, \varphi\right)$, a set of local spherical coordinates over $\man$ adapted to the symmetry.

\subsubsection{Vector fields}
A given vector field $V = V^\mu \partial_\mu$ on $\man$ is cosmologically symmetric if and only if its components in the coordinate basis $\Set{\partial_\mu}$ satisfy
\begin{equation}
\label{eq:cosmo01}
\left(V^\mu\right) = \left(V^t(t), 0, 0, 0\right).
\end{equation}

\subsubsection{$(0,2)$-tensor fields}

Similarly, a given $(0,2)$-tensor field $\tensi{T} = T_{\mu\nu} \diff x^\mu\otimes\diff x^\nu$ is cosmologically symmetric if and only if its components in the coordinate basis $\Set{\partial_\mu}$ satisfy
\begin{equation}
\label{eq:cosmo02}
\left(T_{\mu\nu}\right) = \diag\left(T_{t}(t), \frac{T_{r}(t)}{1 - k r^2}, r^2 T_{r}(t), r^2 \sin(\theta)^2 T_{r}(t)\right),
\end{equation}
where $k \in \Set{-1, 0, 1}$ is the sign of the curvature of the spatial slices. It is worth noting that, in deducing \eqref{eq:cosmo02}, one does not need to assume that $\tensi{T}$ is symmetric; its antisymmetric part vanishes as a consequence of cosmological symmetry.
From this relation, we naturally recover that the components of the most general cosmologically symmetric metric with Lorentz signature $\metric  = g_{\mu\nu} \diff x^\mu\otimes\diff x^\nu$ can be written as
\begin{equation}
\label{eq:cosmog}
\left(g_{\mu\nu}\right) = \diag\left(- N(t)^2, \frac{a(t)^2}{1 - k r^2}, r^2 a(t)^2, r^2 \sin(\theta)^2 a(t)^2\right).
\end{equation}
Of course, in the above, the ``lapse function'' $N(t)$ can be absorbed into the definition of the cosmological time coordinate $t$, so that $g_{00} =-1$. We preferred, yet to leave it general for the sake of flexibility (e.g., if one wants to work in terms of the conformal time $\eta$).

\subsubsection{Differential 1-forms and $(1,1)$-type tensor fields}

Using the musical isomorphisms offered by the metric, \emph{i.e.} if we use the metric to raise or lower indices, we also get that $V_\mu \coloneqq g_{\mu\nu}V^\nu$ and $\ud{T}{\mu}{\nu} \coloneqq g^{\mu\alpha}T_{\alpha\nu}$ must be such that
\begin{equation}
\label{eq:cosmo10}
\left(V_\mu\right) = \left(\theta_V(t), 0, 0, 0\right),
\end{equation}
where we defined $\theta_V(t) \coloneqq -N(t)^2 V^t(t)$, and
\begin{equation}
\label{eq:cosmo11}
\left(\ud{T}{\mu}{\nu}\right) = \diag\left(- \rho_T(t), P_T(t), P_T(t), P_T(t)\right),
\end{equation}
where we defined $\rho_T(t) \coloneqq T_{t}(t)/N(t)^2$ and $P_T(t) \coloneqq T_{r}(t)/a(t)^2$.

Since the metric provides an isomorphism between $TM$ and $T^*M$, we can conclude that \eqref{eq:cosmo10} gives us the most general expression for a differential 1-form and \eqref{eq:cosmo11} the most general expression for a $(1,1)$-tensor field respecting cosmological symmetry; these can also be obtained by direct computation.




\subsection{The $\Lambda$CDM model in a nutshell}

From the theoretical perspective, the $\Lambda$CDM model of cosmology assumes that the universe behaves, at the kinematical level, according to the tools of pseudo-Riemaniann geometry (\emph{i.e.} a metric-affine setup where $\tens{T}\equiv\tens{0}$ \& $\tens{Q}\equiv\tens{0}$) and fixes the dynamics of the universe (the evolution of the scale factor $a(t)$) by means of the following modified Einstein equation \cite{weinberg1972gravitation, 10.1093/lambdaCDMtextbook}
\begin{equation}
\label{eq:LCDM}
\lc{G}_{\mu\nu} + \Lambda g_{\mu\nu} = \kappa T^{\text{(m)}}_{\mu\nu} + \kappa T^{\text{(DM)}}_{\mu\nu},
\end{equation}
where $\lc{\tens{G}} = \lc{G}_{\mu\nu} \diff x^\mu\otimes\diff x^\nu$ is the Einstein tensor, $\Lambda$ is the cosmological constant, $\kappa = 8\pi\mathcal{G}/c^4$, $\tensi{T}^{\text{(m)}} = T^{\text{(m)}}_{\mu\nu} \diff x^\mu\otimes\diff x^\nu$ is the energy-momentum tensor of ordinary matter (including barionic matter and electromagnetic radiation) and $\tensi{T}^{\text{(DM)}} = T^{\text{(DM)}}_{\mu\nu} \diff x^\mu\otimes\diff x^\nu$ is the energy-momentum tensor of dark matter.

Matter is modeled as a perfect fluid
\begin{equation}
\label{eq:perfectfluid}
T^{\text{(m)}}_{\mu\nu} = P g_{\mu\nu} + (\rho+P) u_\mu u_\nu,
\end{equation}
where $g_{\mu\nu}$ is given by \eqref{eq:cosmog}, $\rho$ is the energy density of the fluid, $P$ its pressure and $u = u^\mu \partial_\mu$ is the average 4-velocity of the fluid which is taken to be normalized ($\metric(u, u) = -1$) and such that the fluid is comoving with inertial observers; \emph{i.e.} one has
\begin{equation}
\label{eq:perfectfluid_speed}
\left(u^\mu\right) = \left(1/N(t),0, 0, 0\right).
\end{equation}
To complete this description, properties of matter are encoded by means of an equation of state (EOS) of the form
\begin{equation}
\label{eq:EOS}
P = w \rho,
\end{equation}
where $w\in\mathbb{R}$ is a constant. For instance, for ordinary (dust) matter one has $w_{\text{dust}}=0$ (that is, vanishing pressure), while for electromagnetic radiation (photons), the equation of state is $w_{\text{EM}} = 1/3$ (ensuring that $T^{\text{(EM)}}{}\ud{}{\mu}{\mu} = 0$). 
\bigskip

If we raise the first index of $\tensi{T}^{\text{(m)}}$, we then get
\begin{align}
\left(T^{\text{(m)}}{}\ud{}{\mu}{\nu}\right) &= \diag\left(-\rho(t), P(t), P(t), P(t)\right) \label{eq:tudmatter}\\
                                               &= \rho(t) \diag\left(-1, w, w, w\right).\label{eq:tudmattereos}
\end{align}

In the $\Lambda$CDM model, one considers the dark matter sector to be composed of \emph{cold dark matter}, made of some hypothesized non-relativistic particles, described by a perfect fluid energy-momentum tensor \eqref{eq:perfectfluid} for which $w_{\text{DM}} = 0$. So
\begin{equation}
\label{eq:tudDM}
\left(T^{\text{(DM)}}{}\ud{}{\mu}{\nu}\right) = - \rho_{\text{DM}}(t) \diag\left(1, 0, 0, 0\right).
\end{equation}

The term $\Lambda \metric$ models dark energy. Following the usual reasoning, the term $\kappa \tensi{T}^{\text{(DE)}} \coloneqq -\Lambda \metric$ can be interpreted as a perfect fluid energy-momentum tensor for which $\rho_{\text{DE}} = \Lambda/\kappa$ and $w_{\text{DE}} = -1$, leading to
\begin{equation}
\label{eq:tudDE}
\left(T^{\text{(DE)}}{}\ud{}{\mu}{\nu}\right) = - \rho_{\text{DE}} \diag\left(1, 1, 1, 1\right).
\end{equation}

\subsection{The cosmological smokescreen}\label{sec:observation}

From the computations of section \ref{sec:cosmosymfield}, one can make the following observations  \cite{weinberg1972gravitation}:
\begin{enumerate}
\item Comparing \eqref{eq:perfectfluid_speed} and \eqref{eq:cosmo01} we see that, when required to obey cosmological symmetry, \emph{any} timelike vector field will look like the $4$-velocity of an isotropic observer (and then also like the average $4$-velocity of a perfect fluid).
\item Comparing \eqref{eq:tudmatter} and \eqref{eq:cosmo11} reveals that \emph{any} type $(1,1)$-tensor field which obeys cosmological symmetry must formally look like a perfect fluid energy-momentum tensor.
\end{enumerate}

Obviously, ordinary matter
obeys the above criteria. Assuming cosmological symmetry, this should also be the case for the dark sector energy-momentum tensor (of whatever origin). It then turns out that any cosmologically symmetric extra piece $T^{\text{(D)}}_{\mu\nu}$ of the energy-momentum tensor can be formally split into a ``dark energy'' and a ``dark matter'' part, as seen below:
\begin{equation}
\label{eq:tddDark}
T^{\text{(D)}}_{\mu\nu} \coloneqq T^{\text{(DE)}}_{\mu\nu} + T^{\text{(DM)}}_{\mu\nu}= - \Lambda/\kappa\ g_{\mu\nu} + \rho_{\text{DM}}\ u_\mu u_\nu,
\end{equation}
where $\Lambda$ is, this time, no longer a constant, but a scalar function of $t$.
Indeed, raising the first index, we recover the general expression \eqref{eq:cosmo11} with associated ``density'' $\rho_{\text{D}} \coloneqq \rho_{\text{DM}} + \rho_{\text{DE}} = \rho_{\text{DM}} + \Lambda/\kappa$ and ``pressure'' $P_{\text{D}} \coloneqq  P_{\text{DM}} + P_{\text{DE}}= - \Lambda/\kappa$.
A conservative interpretation of the $\Lambda$CDM model thus \emph{only} indicates that there is a missing piece in Einstein equation $\lc{\tens{G}} = \kappa \tensi{T}^{\text{(m)}}$, but does not allow to draw any stringent conclusion regarding the physical interpretation of the extra piece $\tensi{T}^{\text{(D)}}$. It then looks reasonable to consider the dark sector in full generality, as in \eqref{eq:tddDark}.
The important fact, for us, is that the dark sector energy momentum tensor (of whatever origin) has, in cosmological symmetry, the general form \eqref{eq:tddDark} and one has two degrees of freedom in its choice, represented by the scalars $\Lambda$ and $\rho_{DM}$.

\section{Seed for modified field equations}\label{sec:extendLCDM}

 We will now discuss the construction of an ``educated guess'' for a minimal metric-affine extension of the dynamics of the $\Lambda$CDM model, according to the principles stated in the Introduction. Therefore, for the rest of this paper, we consider the geometric setup as given by a triple $\left(\man, \tens{L}, \metric\right)$ as in Section \ref{sec:setup}.

We start by noting that, in any metric-affine theory of gravity employing the metric and distortion tensor components as dynamical variables, the evolution equations for the metric can be cast into the form
\begin{equation}
\label{eq:idea1}
\lc{\tens{G}} = \kappa \tensi{T}^{\text{(m)}} + \tens{\Theta}^{\text{(D)}},
\end{equation}
where
\begin{equation}
\tens{\Theta}^{\text{(D)}} = \tens{\Theta}^{\text{(D)}}\left(\metric, \partial\metric, \cdots, \partial\cdots\partial\metric; \tens{L}, \partial\tens{L}, \cdots, \partial\cdots\partial\tens{L}\right)
\end{equation}
is a symmetric tensor field of type $(0,2)$ (or $(2,0)$) depending on the metric $\metric$, the distortion tensor components and a finite number of derivatives thereof. Our work will be to determine an ``educated guess'' for $\tens{\Theta}^{\text{(D)}}$.

\vspace{10pt}
\textbf{Step 1.} In the particular case of cosmological symmetry, we want $\tens{\Theta}^{\text{(D)}}$ to geometrically encode the dark sector contributions, in other words, to be the geometric counterpart of $\tensi{T}^{\text{(D)}}$.
As already discussed above, in cosmological symmetry, the most general possible form of $\tens{\Theta}^{\text{(D)}}$ is
\begin{equation}
\label{eq:idea2theta}
\tens{\Theta}^{\text{(D)}} \coloneqq \tens{\Theta}^{\text{(DM)}} + \tens{\Theta}^{\text{(DE)}},
\end{equation}
where $\tens{\Theta}^{\text{(DM)}}$ and $\tens{\Theta}^{\text{(DE)}}$ are interpreted as geometric ``dark matter'' and ``dark energy''.\footnote{This point of view is motivated by the fact that effects attributed to dark-matter are already present at scales smaller than the cosmological one; typically the necessity to introduce dark matter in our physical theory first appeared in the context of rotation curve of galaxies, independently of the presence or not of a cosmological constant (dark-energy) \cite{Zwicky:1933gu}. It is then sound to still model these as distinct physical effects.}
In other words, the following equations of state should hold
\begin{equation}
\label{eq:EOMlambdaCDM}
w_{\text{(DM)}} = 0,\ \ w_{\text{(DE)}} = -1.
\end{equation}
The first relation in \eqref{eq:EOMlambdaCDM} then implies that
\begin{equation}
\label{eq:idea2DMCosmoSym}
\Theta^{\text{(DM)}}_{\mu\nu} \overset{\bullet}{=} V_\mu V_\nu,
\end{equation}
where we use the symbol $\overset{\bullet}{=}$ do denote an equality holding in a cosmologicaly symmetric situation. With the standard $\Lambda$CDM-interpretation, one would have $V_\mu \coloneqq \sqrt{\rho_{\text{(DM)}}} u^\nu g_{\mu\nu}$. Similarly, the second relation in \eqref{eq:EOMlambdaCDM} implies
\begin{equation}
\label{eq:idea2DECosmoSym}
\Theta^{\text{(DE)}}_{\mu\nu} \overset{\bullet}{=} - \Lambda g_{\mu\nu},
\end{equation}
where $\Lambda$ is \emph{a priori} a scalar function (generally, non-constant) and the minus sign is purely conventional.

\bigskip

\textbf{Step 2.} Leaving now the cosmologically symmetric context and going back to the general one, we build a \textit{minimal} metric-affine extension of the $\Lambda$CDM model equations, in the sense that the terms (``degrees of freedom'') describing dark matter and dark energy in a cosmologically symmetric situation are the only ones necessary to desribe the dynamics of these entities in all situations. 
Under this ``minimality'' hypothesis, no extra quantities or terms appear, hence we can simply replace $\overset{\bullet}{=}$ by a strict equality. This gives
\begin{equation}
\label{eq:idea2DM}
\tens{\Theta}^{\text{(DM)}} \coloneqq \underline{V} \otimes \underline{V},
\end{equation}
with $\underline{V} = g_{\mu\nu}V^\mu \diff x^\nu$ for a given vector field $V = V^\mu \partial_\mu$ obtained from the quantities defining the metric-affine geometry; that is $V = V\left(\metric, \partial\metric, \cdots, \partial\cdots\partial\metric; \tens{L}, \partial\tens{L}, \cdots, \partial\cdots\partial\tens{L}\right)$, and
\begin{equation}
\label{eq:idea2DE}
\tens{\Theta}^{\text{(DE)}} \coloneqq - \Lambda \metric,
\end{equation}
where $\Lambda = \Lambda\left(\metric, \partial\metric, \cdots, \partial\cdots\partial\metric; \tens{L}, \partial\tens{L}, \cdots, \partial\cdots\partial\tens{L}\right)$ is a scalar function. 

\vspace{10pt}
To summarize: staying as conservative as possible with respect to the dynamics of the $\Lambda$CDM model, one can consider as our ``educated guess'' an equation of the form
\begin{equation}
\label{eq:idea2}
\lc{\tens{G}} = \kappa \tensi{T}^{\text{(m)}} - \Lambda\metric + \underline{V} \otimes \underline{V},
\end{equation}
where $\Lambda$ is a scalar function and $\underline{V}$ (accordingly, $V$) is a 1-form (accordingly, a vector field) depending on $\metric$, $\tens{L}$ and their derivatives up to a finite order $r$.

Yet, obviously, if we were to stop here, the prescription \eqref{eq:idea2} would be incomplete, for at least two reasons:

\begin{enumerate}
\item In a metric-affine theory of gravity, one has \textit{two} sets of field equations, one for the metric and one for the connection (or the distortion). The above prescription provides an ``educated guess'' for the metric equations only, it makes no claim on the expected form of the connection equation.

\item The above postulated equations are not necessarily variational -- that is, they generally do not arise from an action principle.
\end{enumerate}

A solution to both the above problems is given by an algorithm introduced in \cite{Ducobu:2024grm}, which we present in the next section.

\section{Variational Bootstrapping of $\Lambda$CDM}\label{sec:bootstrap_lambdaCDM}

The variational bootstrapping method \cite{Ducobu:2024grm} allows one to determine a Lagrangian whose Euler-Lagrange equations are the closest to an ``educated guess'' for a given subset of the field equations. After briefly reviewing the algorithm, we apply it to \eqref{eq:idea2}.

\subsection{User's guide on Variational Bootstrapping}\label{sec:userguideVB}

Below, we present a minimalist guide for the method of variational bootstrapping, focusing on computational aspects. For complete and mathematically rigorous details, we refer the reader to \cite{Ducobu:2024grm}.

Assume we want to build a theory involving two distinct sets of dynamical variables, $y^A$ and $z^I$ both depending on the independent variables $x^\mu$.
Generaly speaking, the dynamics of the system must be described by a system of PDEs of an \emph{a priori} given order
\begin{equation}\label{eq:VBtutoeq}
\begin{cases}
\mathcal{Y}_A\left(x^\mu; y^B, \partial_\alpha y^B, \cdots, \partial_{\alpha_1}\cdots\partial_{\alpha_r} y^B; z^J, \partial_\beta z^J, \cdots, \partial_{\beta_1}\cdots\partial_{\beta_s} z^J\right) &= 0\\
\mathcal{Z}_I\left(x^\mu; y^B, \partial_\alpha y^B, \cdots, \partial_{\alpha_1}\cdots\partial_{\alpha_r} y^B; z^J, \partial_\beta z^J, \cdots, \partial_{\beta_1}\cdots\partial_{\beta_s} z^J\right)  &= 0\\
\end{cases}\ \ ,
\end{equation}
with as many $\mathcal{Y}_A$-equations as $y^A$-variables and as many $\mathcal{Z}_I$-equations as $z^I$-variables.

Assume now that we have some insight (an ``educated guess'') regarding the form of the $y^A$-equations only and that this insight is of the form
\begin{equation}\label{eq:VBguess}
\overset{\star}{\mathcal{Y}}_A\left(x^\mu; y^B, \partial_\alpha y^B, \cdots, \partial_{\alpha_1}\cdots\partial_{\alpha_r} y^B; z^J, \partial_\beta z^J, \cdots, \partial_{\beta_1}\cdots\partial_{\beta_s} z^J\right) = 0,
\end{equation}
with as many equations as $y^A$-variables.\\
Variational bootstrapping then allows one to find a canonical correction term to be added to \eqref{eq:VBguess} in such a way that the corrected $y^A$-equations are derived from a Lagrangian; moreover, this Lagrangian is uniquely determined up to boundary terms and to terms that do not involve $y^A$. This will allow us, under certain circumstances (to be detailed below) to also ``bootstrap'' the $z^I$-equation.
The Lagrangian density of the said Lagrangian is given\footnote{In principle, such a Lagrangian can only be constructed in a coordinate chart. Yet, if our variables $y^A$ are tensor ones and the left hand sides of \eqref{eq:VBguess} are tensor densities, then the Lagrangian is globally well defined -- and actually, generally covariant.} by
\begin{equation}
\label{eq:VBlag}
\lagrangian_{y} = y^A \int_{0}^{1} \overset{\star}{\mathcal{Y}}_A\left(x^\mu; u y^B, u \partial_\alpha y^B, \cdots, u \partial_{\alpha_1}\cdots\partial_{\alpha_r} y^B; z^J, \partial_\beta z^J, \cdots, \partial_{\beta_1}\cdots\partial_{\beta_s} z^J\right) \dint u,
\end{equation}
where the $y^B$ variables and all their derivatives are scaled by the same $u$ factor in the integrand and a sum over index $A$ is understood.

In cases where our guessed equations \eqref{eq:VBguess} can be obtained from a variational principle, any Lagrangian density producing \eqref{eq:VBguess} as $y^A$-part of the Euler-Lagrange equations must be of the form \eqref{eq:VBlag} up to Lagrangian densities completely independent of the $y^A$-variables (\emph{i.e.} which do not contribute to the $y^A$-equations) and total divergences (which do not contribute to the field equations at all), see \cite{Ducobu:2024grm}.
The form of the $y^A$-independent terms in the Lagrangian density is not constrained by the above procedure and should thus be found by different means\footnote{Intuitively, the situation is similar to try finding a function $f : \mathbb{R}^2 \to \mathbb{R}^2 : (x,y) \mapsto f(x,y)$ when the only information we have is on $\partial_x f$.}.

But what if the equation \eqref{eq:VBguess} we start with is not variational (\emph{i.e.} what if it does not arise as the Euler-Lagrange equation with respect to  $y^A$ of any Lagrangian)? In that case, the idea behind variational bootstrapping is to still compute the Lagrangian density \eqref{eq:VBlag}. This does two things for us:

\begin{enumerate}
\item It provides a Lagrangian density that will ``correct the $y^A$-equations'' into their ``closest'' variational equations.\footnote{To be more precise, the difference between \eqref{eq:VBguess} and the $y^A$-part of the Euler-Lagrange equations associated with \eqref{eq:VBlag} is a linear combination of the components of the so-called Helmholtz form, which precisely measures the obstruction from variationality of \eqref{eq:VBguess}.},

\item It constrains the acceptable form (up to $y^A$-independent terms) for the $z^I$-equations.
\end{enumerate}
In that case, the $y^A$-part of the Euler-Lagrange equations associated with \eqref{eq:VBlag} will differ from \eqref{eq:VBguess} but \eqref{eq:VBlag} will provide a canonical (and ``minimal'') Lagrangian based on the guess \eqref{eq:VBguess}.

\vspace{20pt}
In the case of modified gravity theories based on GR, the field variables are usually the metric $\metric$ and, generally speaking, another set of tensor variables, say $\tensi{\psi}^A$. Our knowledge on metric-based approaches to gravity provides us with an ``educated guess'' for the metric equation, but not necessarily for the dynamics  of the $\tensi{\psi}^A$ fields.\footnote{One might have an idea of the form of the dynamical equation for the $\tensi{\psi}^A$ variables if those fields have some known dynamics (\emph{e.g.,} on Minkowski spacetime), which we try to extend in the presence of a generic metric (think \emph{e.g.} to a scalar field). In such a case, we verified that the variational bootstrapping procedure was in line with the usual approaches, see \cite{Ducobu:2024grm}.} One can then apply variational bootstrapping with respect to the metric to obtain a Lagrangian density based on this guessed equation and then deduce the $\tensi{\psi}^A$ dynamics from it.

In this case, assuming our guess is schematically of the form
\begin{equation}\label{eq:initialguess}
\lc{G}^{\mu\nu} = T^{\mu\nu}\left(\metric, \partial\metric; \tensi{\psi}^A, \partial\tensi{\psi}^A\right),
\end{equation}
where $T^{\mu\nu}\left(\metric, \partial\metric; \tensi{\psi}^A, \partial\tensi{\psi}^A\right)$ stands for the energy-momentum tensor associated with the extra fields $\tensi{\psi}^A$, the procedure prescribes to compute\footnote{where the lower integration bound is here understood as a limit, see \cite{Ducobu:2024grm}.}
\begin{equation}\label{eq:vbmetric}
\begin{split}
  \lagrangian_{g} & = g_{\mu\nu} \int_0^1 \lc{G}^{\mu\nu}\left(u \metric, u \partial\metric, u \partial\partial\metric\right) \sqrt{-\det(ug)}\ \dint u\\
                  &- g_{\mu\nu} \int_0^1 T^{\mu\nu}\left(u \metric, u \partial\metric; \tensi{\psi}^A, \partial\tensi{\psi}^A\right) \sqrt{-\det(ug)}\ \dint u,\\
                & \eqqcolon \lagrangian_{\text{\tiny EH}} + \lagrangian_\psi.\\
\end{split}
\end{equation}

Here again, this procedure provides (up to boundary terms)  the ``best'' Lagrangian density related to our initial guess \eqref{eq:initialguess}, up to terms that would not give any contribution to the metric equation -- that is, Lagrangian densities built without using the metric $\metric$ at all. Such Lagrangian densities are quite uncommon for physical theories -- as one usually uses the metric in the construction of the Lagrangian density to manipulate indices and/or to define a volume element. They are, nevertheless, not ruled out \emph{a priori} by the above construction. In \cite{Ducobu:2024grm}, we classified all possible 4-dimensional natural Lagrangians that are independent of the metric in the context of a metric-affine setup. It turns out there is a limited amount of these which are obtained from specific $4$-forms constructed solely from the components of the distorsion tensor $\tens{L}$ and their first derivatives, see \cite{Ducobu:2024grm} for a complete classification.

\subsection{General requirements}\label{sec:hypothesis}

Let us now apply the above algorithm to our ``educated guess'' \eqref{eq:idea2}. 
To proceed with the construction of the Lagrangian density, we first need to refine the requirements on our field equations and on the quantities $\Lambda$ and $V$ appearing in \eqref{eq:idea2DM}--\eqref{eq:idea2DE}. In what follows, we make the three following assumptions:

\begin{enumerate}
\item\label{hyp:depgandL} The quantities $\Lambda$ and $V$ must depend on $\metric$, $\tens{L}$, and their first order derivatives $\partial\metric$, $\partial\tens{L}$ only and in a generaly covariant way, more precisely:\footnote{This hypothesis does not contradict the fact that our expression could depend on $\nabla \tens{L}$, since \[\nabla_\alpha \ud{L}{\rho}{\mu\nu} = \lc{\nabla}_\alpha \ud{L}{\rho}{\mu\nu} + \ud{L}{\rho}{\beta\alpha} \ud{L}{\beta}{\mu\nu} - \ud{L}{\beta}{\mu\alpha} \ud{L}{\rho}{\beta\nu} - \ud{L}{\beta}{\nu\alpha} \ud{L}{\rho}{\mu\beta}.\] We can then write $\Lambda = \Lambda(\metric,\tens{L}, \lc{\nabla}\tens{L})$ without loss of generality.}
\begin{equation} \label{eq:dep_g_L}
\Lambda = \Lambda(\metric, \tens{L}, \lc{\nabla}\tens{L}),\,\,
V = V(\metric, \tens{L}, \lc{\nabla}\tens{L}).
\end{equation}
In particular, this ensures that the guessed equations \eqref{eq:idea2} are generally covariant and that the Euler-Lagrange equations of the Lagrangian obtained by variational bootstrapping are at most of second order in  both $\metric$ and $\tens{L}$.

\item\label{hyp:useg} In densitizing the seed equations (``educated guess''), the volume form used is the Riemannian one $\tensi{\epsilon}_g=\sqrt{-\det(g)}\diff x$ and indices are lowered/raised by $\metric$.\\
An immediate consequence of this assumption is that \textit{all} Lagrangian terms will exhibit a nontrivial dependence in $\metric$ -- and hence the full Lagrangian can be recovered, up to boundary terms, by variational boostrapping with respect to the metric.

\item\label{hyp:homogeneity} We can write $\Lambda$ as a finite sum \[\Lambda = \sum_k \Lambda_{(k)},\] were each $\Lambda_{(k)}$ is a homogeneous function of degree $k$ in $\metric$ and its derivatives; and the same property holds for $V$.
\end{enumerate}

\subsection{Bootstrapping of $\Lambda$CDM}\label{sec:VBLCDM}

We want to construct here the Lagrangian density obtained by variational bootstrapping of \eqref{eq:idea2}. In the following, we will apply the variational bootstrapping procedure to the gravity+dark sector only. To include regular matter, one should just add the corresponding matter Lagrangian to the one obtained below.
According to \eqref{eq:vbmetric} and the above hypotheses, we thus need to compute
\begin{equation}
\label{eq:Lg}
\lagrangian_{g} = \lagrangian_{\text{\tiny EH}} + \lagrangian_{\Lambda} + \lagrangian_{V},
\end{equation}
where
\begin{equation}
\label{eq:LEHint}
\lagrangian_{\text{\tiny EH}} = g_{\mu\nu} \int_0^1 \lc{G}^{\mu\nu}\left(u \metric, u \partial\metric, u \partial\partial\metric\right) \sqrt{-\det(ug)}\ \dint u,
\end{equation}
and\footnote{Note -- see Appendix \ref{sec:identity} -- that under the rescaling $g_{\mu\nu} \mapsto u g_{\mu\nu}$, the Christoffel symbols $\lc{\Gamma}^{\rho}_{\mu\nu}$ remain invariant.}

\begin{equation}
\label{eq:LL}
\lagrangian_{\Lambda} = g_{\mu\nu} \int_0^1 \Lambda(u\metric, \tens{L}, \lc{\nabla}\tens{L}) u^{-1} g^{\mu\nu}\sqrt{-\det(ug)}\ \dint u,
\end{equation}

\begin{equation}
\label{eq:LV}
\lagrangian_{V} = - g_{\mu\nu} \int_0^1 V^\mu(u\metric, \tens{L}, \lc{\nabla}\tens{L})V^\nu(u\metric, \tens{L}, \lc{\nabla}\tens{L})\sqrt{-\det(ug)}\ \dint u.
\end{equation}

Using the relations from Appendix \ref{sec:identity}, following the usual computation, we get that

\begin{equation}
\label{eq:LEH}
\lagrangian_{\text{\tiny EH}} = g_{\mu\nu}\lc{G}^{\mu\nu}\sqrt{-\lvert g\rvert} = - \lc{R} \sqrt{-\lvert g \rvert},
\end{equation}
with $\lvert g\rvert \coloneqq \det(g)$.

From equation \eqref{eq:LL}, we get

\begin{equation}
\label{eq:LL2}
\lagrangian_{\Lambda} = 4 \sqrt{-\lvert g\rvert} \int_0^1 u\ \Lambda(u\metric,\tens{L}, \lc{\nabla}\tens{L})\ \dint u.
\end{equation}

Under our hypothesis \ref{hyp:homogeneity}, we can explicitly compute $\lagrangian_{\Lambda}$; to this aim, denote
\begin{equation}
\label{eq:LLk}
\Lambda(\metric, \tens{L}, \lc{\nabla}\tens{L}) = \sum_{k=M}^N \Lambda_{(k)}(\metric, \tens{L}, \lc{\nabla}\tens{L}),
\end{equation}
where $M,N\in \mathbb{Z}$ and each $\Lambda_{(k)}$ is homogneous of degree $k$ in $\metric$, that is:
\begin{equation}
\Lambda_{(k)}(u\metric,\tens{L}, \lc{\nabla}\tens{L}) = u^k \Lambda_{(k)}(\metric, \tens{L}, \lc{\nabla}\tens{L}).
\end{equation}

In this case, \eqref{eq:LL2} becomes
\begin{equation}
\label{eq:LL3}
\lagrangian_{\Lambda} = 4 \sqrt{-\lvert g\rvert} \sum_{k=M}^N \frac{\Lambda_{(k)}(\metric, \tens{L}, \lc{\nabla}\tens{L})}{k+2}.
\end{equation}

Following the same line, \eqref{eq:LV} can be written as
\begin{equation}
\label{eq:LV2}
\lagrangian_{V} = - g_{\mu\nu} \sqrt{-\lvert g\rvert} \int_0^1 u^2\ V^\mu(u\metric, \tens{L}, \lc{\nabla}\tens{L})V^\nu(u\metric, \tens{L}, \lc{\nabla}\tens{L})\ \dint u.
\end{equation}

Here again, according to hypothesis \ref{hyp:homogeneity}, we want to rewrite $V^\mu(\metric, \tens{L}, \lc{\nabla}\tens{L})$ as 
\begin{equation}
\label{eq:LVk}
V^\mu(\metric, \tens{L}, \lc{\nabla}\tens{L}) = \sum_{k=I}^J V^\mu_{(k)}(\metric, \tens{L}, \lc{\nabla}\tens{L}),
\end{equation}
where $I,J\in \mathbb{Z}$ with the defining property that
\begin{equation}
V^\mu_{(k)}(u\metric,\tens{L}, \lc{\nabla}\tens{L}) = u^k V^\mu_{(k)}(\metric, \tens{L}, \lc{\nabla}\tens{L}).
\end{equation}

Using \eqref{eq:LVk}, \eqref{eq:LV2} gives
\begin{equation}
\label{eq:LV3}
\lagrangian_{V} = - \sqrt{-\lvert g\rvert} g_{\mu\nu} \sum_{k=I}^J\sum_{l=I}^J \frac{V^\mu_{(k)}(\metric, \tens{L}, \lc{\nabla}\tens{L})V^\nu_{(l)}(\metric, \tens{L}, \lc{\nabla}\tens{L})}{k+l+3}.
\end{equation}

\subsection{Variationally completed field equations}\label{sec:fieldeqVBLCDM}

In order to determine the field equations associated to the Lagrangian density \eqref{eq:Lg}, we will separately compute the variation of each one of the terms in the summations of \eqref{eq:LL3} and \eqref{eq:LV3}.

\subsubsection{Variation with respect to $\metric$}
Using the relations from Appendix \ref{sec:identity}, as expected, the variation of \eqref{eq:LEH} will give
\begin{equation}
\label{eq:varLEH}
\delta_g\lagrangian_{\text{\tiny EH}} = \lc{G}^{\mu\nu}\delta g_{\mu\nu} \sqrt{-\lvert g\rvert} +  \lc{\nabla}_\mu\left(g^{\alpha\mu}\delta_g\lc{\Gamma}^{\beta}_{\alpha\beta} - g^{\alpha\beta}\delta_g\lc{\Gamma}^{\mu}_{\alpha\beta} \right) \sqrt{-\lvert g\rvert} \simeq \sqrt{-\lvert g\rvert} \lc{G}^{\mu\nu}\delta g_{\mu\nu},
\end{equation}
where $\simeq$ means equality up to boundary terms.
Let us now come to the variation of \eqref{eq:LL3} with respect to the metric. To this purpose, we calculate separately the variation of each term $\Lambda_{(k)}$. Using relation \eqref{eq:vardetg} in Appendix \ref{sec:identity}, we get:
\begin{equation}
\label{eq:varLLk}
\delta_g\left(\Lambda_{(k)} \sqrt{-\lvert g\rvert}\right) = \left(\frac{\partial \Lambda_{(k)}}{\partial g_{\mu\nu}}\delta g_{\mu\nu} + \left(\partial\Lambda_{(k)}\right)^{\alpha\phantom{\rho}\sigma\beta}_{\phantom{\alpha}\rho}\ \delta_g\left(\lc{\nabla}_\alpha \ud{L}{\rho}{\sigma\beta}\right)\right)\sqrt{-\lvert g\rvert} + \frac{1}{2} \Lambda_{(k)} g^{\mu\nu} \delta g_{\mu\nu} \sqrt{-\lvert g\rvert},
\end{equation}
where, to shorten the notation, we have defined $\left(\partial\Lambda_{(k)}\right)^{\alpha\phantom{\rho}\sigma\beta}_{\phantom{\alpha}\rho} \coloneqq \frac{\partial \Lambda_{(k)}}{\partial\left(\lc{\nabla}_\alpha \ud{L}{\rho}{\sigma\beta}\right)}$. Using again the relations from Appendix \ref{sec:identity} (in particular \eqref{eq:varGamma} and \eqref{eq:gvarL}), we get that\footnote{Along the computation, it is also particularly useful to use the identity \[A^{\mu\nu\alpha} \left(B_{\mu\nu\alpha} - B_{\alpha\mu\nu} + B_{\nu\alpha\mu}\right) = \left(A^{\mu\nu\alpha} - A^{\alpha\mu\nu} + A^{\nu\alpha\mu}\right) B_{\nu\alpha\mu}
  \] valid for any two objects with three indices.}
  
\begin{equation}
\label{eq:deltaLLk}
\delta_g\left(\Lambda_{(k)} \sqrt{-\lvert g\rvert}\right) \simeq \sqrt{-\lvert g\rvert}\ T^{\mu\nu}_{(k)}\ \delta g_{\mu\nu} ,
\end{equation}
with
\begin{eqnarray} \label{eq:TLk}
	T_{\left(k\right)}^{\mu \nu } &=&\frac{\partial \Lambda _{\left(k\right)}}{\partial g_{\mu \nu }}+\frac{1}{2}\Lambda _{\left( k\right)}g^{\mu \nu } \\
	&&+\frac{1}{2}\lc{\nabla}_{\alpha}\left\{ \left( \partial \Lambda _{\left( k\right) }\right) _{~\rho }^{\tau ~\sigma \beta }\left( -L_{~\sigma\beta}^{\gamma}g^{\rho\xi}B_{\xi\gamma\tau}^{\mu\nu\alpha} + L_{~~\beta}^{\rho \gamma}B_{\gamma\sigma\tau}^{\mu\nu\alpha} + L_{~\sigma}^{\rho ~\gamma}B_{\gamma\beta\tau}^{\mu\nu\alpha}\right) \right\}   \notag
\end{eqnarray}%
where $B_{\xi \gamma \tau }^{\mu \nu \alpha }$ is a combination of Kronecker deltas: 
\begin{equation}
B_{\xi \gamma \tau }^{\mu \nu \alpha }:=\frac{1}{2}\left( \delta_{\xi}^{\mu}\delta_{\gamma}^{\nu}\delta_{\tau}^{\alpha} + \delta_{\xi}^{\nu}\delta_{\gamma}^{\mu}\delta_{\tau}^{\alpha} - \delta_{\tau}^{\mu}\delta_{\gamma}^{\nu}\delta_{\xi}^{\alpha} - \delta_{\tau}^{\nu}\delta_{\gamma}^{\mu}\delta_{\xi}^{\alpha} + \delta_{\xi}^{\mu}\delta_{\tau}^{\nu}\delta_{\gamma}^{\alpha} + \delta_{\xi}^{\nu}\delta_{\tau}^{\mu}\delta_{\gamma}^{\alpha}\right) .
\end{equation}


To compute the variation of $\lagrangian_V$ with respect to the metric, we can reuse the results of the previous computation. Indeed, if we introduce
\begin{equation}
\label{eq:Vkl}
V_{(kl)} \coloneqq g_{\mu\nu}V^\mu_{(k)} V^\nu_{(l)},
\end{equation}
we can write that
\begin{equation}
\label{eq:deltaLVkl}
\delta_g\left(V_{(kl)} \sqrt{-\lvert g\rvert}\right) \simeq \sqrt{-\lvert g\rvert}\ T^{\mu\nu}_{(kl)}\ \delta g_{\mu\nu},
\end{equation}
where the expression for $T^{\mu\nu}_{(kl)}$ is obtained by performing the replacement $\Lambda_{(k)} \to V_{(kl)}$ in \eqref{eq:TLk}.

Combining \eqref{eq:varLEH}, \eqref{eq:deltaLLk} and \eqref{eq:deltaLVkl}, we get that the metric field equation for the Lagrangian \eqref{eq:Lg}, in the absence of matter source, is
\begin{equation}
\label{eq:EELmetric}
\lc{\tens{G}} = - 4 \sum_{k=M}^{N} \frac{1}{k+2} \tensi{T}^{(k)} + \sum_{k=I}^{J}\sum_{l=I}^{J} \frac{1}{k+l+3} \tensi{T}^{(kl)};
\end{equation}
where $\tensi{T}^{(k)} = T^{(k)}_{\mu\nu}\diff x^\mu\otimes\diff x^\nu$ and $\tensi{T}^{(kl)} = T^{(kl)}_{\mu\nu}\diff x^\mu\otimes\diff x^\nu$ are as above.


\subsubsection{Variation with respect to $\tens{L}$}

As already mentioned, even though our initial guess was only related to the metric equation, variational bootstrapping allows us to completely determine the Lagrangian, up to boundary terms\footnote{Remember that our hypothesis \ref{hyp:useg} excludes Lagrangians independent of $\metric$.}. In particular the distortion field equation is also completely determined.

Let us now compute these field equations. Since $\lagrangian_{\text{\tiny EH}}$ is independent of $\tens{L}$, only $\lagrangian_{\Lambda}$ and $\lagrangian_{V}$ will contribute.

For $\lagrangian_{\Lambda}$, we find
\begin{equation}
\delta_L\left(\Lambda_{(k)}\sqrt{-\lvert g\rvert}\right) = \delta_L\Lambda_{(k)}\ \sqrt{-\lvert g\rvert} = \left(\frac{\partial \Lambda_{(k)}}{\partial \ud{L}{\rho}{\sigma\beta}}\delta \ud{L}{\rho}{\sigma\beta} + \left(\partial\Lambda_{(k)}\right)^{\alpha\phantom{\rho}\sigma\beta}_{\phantom{\alpha}\rho}\ \delta_L\left(\lc{\nabla}_\alpha \ud{L}{\rho}{\sigma\beta}\right)\right)\sqrt{-\lvert g\rvert}.
\end{equation}
Using identity \eqref{eq:varL} and integration by parts, we can write
\begin{equation}
\label{eq:LvarLLk}
\delta_L\left(\Lambda_{(k)}\sqrt{-\lvert g\rvert}\right) \simeq  \sqrt{-\lvert g\rvert}\ \left(\Psi_{(k)}\right)\du{}{\rho}{\sigma\beta}\ \delta \ud{L}{\rho}{\sigma\beta},
\end{equation}
where
\begin{equation}
\label{eq:psiLk}
\left(\Psi_{(k)}\right)\du{}{\rho}{\sigma\beta} \coloneqq \frac{\partial\Lambda_{(k)}}{\partial\ud{L}{\rho}{\sigma\beta}} - \lc{\nabla}_\alpha\left[\left(\partial\Lambda_{(k)}\right)^{\alpha\phantom{\rho}\sigma\beta}_{\phantom{\alpha}\rho}\right].
\end{equation}

Here again, for $\lagrangian_{V}$, one can reuse the previous result by introducing $V_{(kl)}$ from \eqref{eq:Vkl} to obtain that
\begin{equation}
\label{eq:LvarLVk}
\delta_L\left(V_{(kl)}\sqrt{-\lvert g\rvert}\right) \simeq  \sqrt{-\lvert g\rvert}\ \left(\Psi_{(kl)}\right)\du{}{\rho}{\sigma\beta}\ \delta \ud{L}{\rho}{\sigma\beta},
\end{equation}
where the expression of $\left(\Psi_{(kl)}\right)\du{}{\rho}{\sigma\beta}$ is obtained via the replacement $\Lambda_{(k)} \to V_{(kl)}$ in \eqref{eq:psiLk}.

This finally gives us a distortion field equation of the form
\begin{equation}
\label{eq:EELdistortion}
4 \sum_{k=M}^{N} \frac{1}{k+2} \tens{\Psi}_{(k)} - \sum_{k=I}^{J}\sum_{l=I}^{J} \frac{1}{k+l+3} \tens{\Psi}_{(kl)} = 0,
\end{equation}
where $\tens{\Psi}_{(k)} \coloneqq \left(\Psi_{(k)}\right)\du{}{\rho}{\sigma\beta} \diff x^\rho \otimes \partial_\sigma \otimes \partial_\beta$ and $\tens{\Psi}_{(kl)} \coloneqq \left(\Psi_{(kl)}\right)\du{}{\rho}{\sigma\beta} \diff x^\rho \otimes \partial_\sigma \otimes \partial_\beta$.

\subsection{Analysis of the equations}\label{sec:Analysiseq}

Before closing our discussion, let us comment on the level of generality of the obtained expressions and on how these relates to previously formulated models in metric-affine gravity. Because of the freedom of choice in $\Lambda$ and $V$, our Lagrangian \eqref{eq:Lg} still encompasses many of the models present in the literature.
For a state of the art on metric-affine theories of gravity, we redirect the reader to \cite{CANTATA:2021ktz} while, for a review on historical developments, one can see \cite{Blagojevic:2012bc, Puetzfeld:2004yg}.

The most straightforward example is that of a metric-affine theory whose Lagrangian is the Ricci scalar $R$ of the independent connection $\Gamma$, see \emph{e.g.} \cite{Shimada:2018lnm, Iosifidis:2020gth}. Because of the well-known geometric identity
\begin{equation}
\label{eq:Ridentity}
R \equiv \lc{R} + 2 \left(\ud{L}{\mu}{\nu\left[\mu\right\vert}\ud{L}{\nu\rho}{\left\vert\rho\right]} + \lc{\nabla}_{\left[\mu\right\vert} \ud{L}{\mu\nu}{\left\vert\nu\right]}\right),
\end{equation}
the choice\footnote{where each term is $(-1)$-homogeneous with respect to $\metric$}
\begin{equation}
\label{eq:LambdaRidentity}
\Lambda(\metric, \tens{L}, \lc{\nabla}\tens{L}) = - \frac{1}{2} \left(\ud{L}{\mu}{\nu\left[\mu\right\vert}\ud{L}{\nu\rho}{\left\vert\rho\right]} + \lc{\nabla}_{\left[\mu\right\vert} \ud{L}{\mu\nu}{\left\vert\nu\right]}\right)
\end{equation}
and $V(\metric, \tens{L}, \lc{\nabla}\tens{L}) = 0$ in the seed equation \eqref{eq:idea2} will lead, after variational bootstrapping, to a Lagrangian given by the right-hand side of \eqref{eq:Ridentity}.

A more interesting example is that of quadratic metric-affine theories \cite{PhysRevD.56.7769, Vitagliano:2010sr}, \emph{i.e.} theories whose Lagrangian contains terms that are quadratic in the curvature $\tens{R}$ and/or torsion $\tens{T}$ and/or non-metricity $\tens{Q}$ of $\Gamma$. In this case, one can easily see that appropriate choices of $\Lambda$ and $V$ can generate all the terms under consideration in \cite{PhysRevD.56.7769} or \cite{Vitagliano:2010sr} \emph{except} those which are quadratic in $\tens{R}$ (\emph{e.g.} the term proportional to the coefficient $z_4$ in equation (3.1) of \cite{PhysRevD.56.7769}). This comes from our hypothesis \ref{hyp:depgandL} on the form of $\Lambda$ and $V$, which implies that our Lagrangians can only contain a linear term in $\lc{R}$.

Following the same line, our procedure then also selects a restrictive class of quadratic gauge theories of gravity within a metric-affine framework as it constrains the dependence in $\tens{R}$ (Compare with chapter 9 of \cite{Blagojevic:2012bc}).

Finally, let us mention that, while we here considered theories based on a fully general metric-affine setup, many models studied in the literature start with some restriction on the connection $\Gamma$; typically $\tens{Q}\equiv\tens{0}$, as in Einstein-Cartan gravity or Poincaré gauge gravity, see respectively chapters 4 and 5 of \cite{Blagojevic:2012bc}. Naturally, this would not affect the behavior of our procedure and the conclusion would remain the same\footnote{Technically, imposing restrictions on the connection at the kinematical level account for changing the configuration manifold on which our Lagrangian theory is based.
For example, by imposing $\tens{Q}\equiv\tens{0}$ at the kinematical level, one assumes the configuration manifold to be
\[Y = \met(\man)\times_{\man} \tor(\man),\]
where $\tor(\man)$ denotes the bundle of torsions -- \emph{i.e.} the bundle of type $(1,2)$ tensor fields antisymmetric over their two lower indices -- instead of
\[Y = \met(\man)\times_{\man} T^1_2\man\]
as in this paper, see \cite{Ducobu:2024grm}. Nevertheless, since we are interested in variational bootstrapping with respect to the metric anyway, the computation and conclusions would perform exactly the same as in our case.}: Variational bootstrapping of the $\Lambda$CDM equations is compatible with quadratic metric-affine theories of gravity \emph{except} for terms that are non-linear in the curvature tensor $\tens{R}$.

\vspace{10pt}
As a final comment, let us emphasize that the particularity of the method outlined here in this business of finding Lagrangians is that it directly relates the notion of ``minimality'' to the variational structure of the theory. In other words, variational bootstrapping allows us to select metric-affine models which are minimal extensions of the $\Lambda$CDM model \emph{in a variational sense} and this, as we just saw, allows for a more restrictive selection of Lagrangians.

\section{Conclusion}\label{sec:ccl}

In this article, we studied the application of variational bootstrapping, first discussed in \cite{Ducobu:2024grm}, to metric-affine theories of gravity.
Through a careful examination of the $\Lambda$CDM model of cosmology, we proposed a minimal modification of GR's field equation, based on metric-affine geometry; our hypotheses for that construction are explained in section \ref{sec:extendLCDM} (see also subsection \ref{sec:hypothesis}).
Then, variational bootstrapping allowed us to systematically construct the Lagrangian for the full theory, leading to the ``corrected'' metric equations \eqref{eq:EELmetric} and, at the same time, completely determining the field equations for the distortion tensor (independent connection) \eqref{eq:EELdistortion}.

We showed in subsection \ref{sec:Analysiseq} how our model (which is actually, a whole family of metric-affine models, obeying certain minimality criteria) relates to previously formulated ones.
We saw that the notion of ``variationaly minimal extension'' outlined in this paper gives a more restrictive criterion to select Lagrangians than the ones usually considered in the literature when considering \emph{e.g.,} quadratic metric-affine theories. Naturally, this does not rule out the other existing models \emph{per se}! It should nevertheless be regarded as an interesting property since, in modified gravity in general (and in metric-affine gauge theories of gravity in particular), one usually struggles to single out a theory exhibiting a desired behavior.

Last but not least, this investigation demonstrates that variational bootstrapping is a powerful tool to construct theoretical models based on phenomenological requirements. The most remarkable feature of the method is that it allows to obtain a sensible Lagrangian for the theory under study by only guessing \emph{a part of} the system's dynamics.

\vspace{10pt}
An important direction to extend the present work is to investigate solutions of the field equations \eqref{eq:EELmetric}$\land$\eqref{eq:EELdistortion}. In particular, it would be interesting to see if a well-chosen form of the scalar function $\Lambda$ and vector field $V$ would allow, in a cosmological setup, to dynamically produce a behavior of the scale factor $a(t)$ compatible with the current universe accelerated expansion and/or with inflation.

In addition to that, it would also be necessary to study the behavior of compact objects (black holes, neutron stars) in our theory to see how these differ from the GR ones -- and to which extent this remains compatible with modern observations.

Another possible extension includes the application of our method to metric-affine theories using a tetrad and a spin connection as dynamical variables and the study of the conditions under which this approach is equivalent to the one presented here.

We leave these and further questions for future works.

\section*{Acknowledgements}
L.D. acknowledges support from the Transilvania Fellowship Program for Postdoctoral Research/Young Researchers (September 2022) and from a Postdoctoral sojourn grant from \textsc{Complexys} Institute from University of Mons (Belgium) and also acknowledges networking support by the COST Action CA18108 in the early developments of the present project.
N.V. acknowledges networking support by the COST Action 21136.

\bibliographystyle{utphys}
\bibliography{references}

\appendix
\section{Cosmological symmetry}\label{sec:cosmosym}

\subsection{Symmetry}

As discussed in \cite{Hohmann:2019fvf}, when saying that a given tensor field $\tensi{T}$ presents a symmetry, one means that our manifold is equipped with a certain group action
\[\Phi : G \times \man \to \man : (g, x) \mapsto \Phi(g,x) \eqqcolon \Phi_g(x),\]
smooth over its second variable, and that $\tensi{T}$ is invariant under this group action in the sense that
\begin{equation}
\label{eq:sym}
\forall g\in G,\ \Phi_g^*(\tensi{T}) = \tensi{T}.
\end{equation}

When $G$ is a Lie group, it is usually more convenient to study this notion of symmetry in terms of its infinitesimal generators. Let us denote by $\mathfrak{g}$ the Lie algebra of $G$ and by $X_\xi$ the generating vector field (on $M$), corresponding to a given $\xi\in\mathfrak{g}$\footnote{Technically, this means that $X_\xi$ is the vector field on $M$, whose integral curves are given by the flow of $\Phi$, \emph{i.e.}, at any point $p\in\man$, the integral curve of $X_\xi$ through $p$ is \[\curve^p_{X_\xi} : t \mapsto \Phi(\exp(t\xi),p),\] where $\exp: \mathfrak{g} \to G$ is the exponential map.}. If a tensor field $\tensi{T}$ is invariant under the action of $G$, then
\begin{equation}
\label{eq:Liesym}
\forall \xi\in \mathfrak{g},\ \Lie_{X_\xi}{\tensi{T}} = \tensi{0},
\end{equation}
where $\Lie$ denotes the Lie derivative.

Moreover, if $G$ is connected, relations \eqref{eq:sym} and \eqref{eq:Liesym} are actually, equivalent and one can then derive the conditions for a tensor field $\tensi{T}$ to present a given symmetry via \eqref{eq:Liesym}. It is the condition that we will use in the following.

Using this definition of symmetry, a metric-affine spacetime will be symmetric under a given group $G$ provided both the metric $\metric$ and the distortion $\tens{L}$ are invariant under the action of $G$.

\subsection{Generators of cosmological symmetry}\label{sec:cosmosymvec}

Cosmological symmetry refers to the idea that, following the cosmological principle, no observer has a prefered place in the universe, so that spacetime must be spatially homogeneous and isotropic. More precisely here, we then expect our setup to be invariant under both the ``spatial'' rotation group $SO(3)$ (providing isotropy) and a group representing spatial (quasi-)translations (providing spatial homogeneity).

 One thus obtains 6 generators of cosmological symmetry $X_i$ $(i = 1, \cdots, 6)$. In local spherical coordinates $ \left(x^\mu\right) = \left(t, r, \theta, \varphi\right)$ on $\man$, $SO(3)$-generators are described as:
\begin{equation}
\label{eq:gen_rot}
X_{1} = \sin(\varphi) \partial_\theta + \frac{\cos(\varphi)}{\tan(\theta)} \partial_\varphi,\ X_{2} = - \cos(\varphi) \partial_\theta + \frac{\sin(\varphi)}{\tan(\theta)} \partial_\varphi,\ X_{3} = - \partial_\varphi,
\end{equation}
as they satisfy the commutation relation of $\mathfrak{so}(3)$. 
Generators of (quasi-)translations are, \cite{weinberg1972gravitation},
\begin{align}
\label{eq:gen_trans}
  X_{4} = &\sqrt{1 - k r^2} \left(\sin(\theta)\cos(\varphi)\partial_r + \frac{1}{r} \cos(\theta)\cos(\varphi)\partial_\theta - \frac{1}{r} \frac{\sin(\varphi)}{\sin(\theta)}\partial_\varphi\right),\\
  X_{5} = &\sqrt{1 - k r^2} \left(\sin(\theta)\sin(\varphi)\partial_r + \frac{1}{r} \cos(\theta)\sin(\varphi)\partial_\theta + \frac{1}{r} \frac{\cos(\varphi)}{\sin(\theta)}\partial_\varphi\right),\\
  X_{6} = &\sqrt{1 - k r^2} \left(\cos(\theta)\partial_r - \frac{1}{r} \sin(\theta)\partial_\theta\right),
\end{align}
where $k \in \Set{-1, 0, 1}$ is the sign of the spatial curvature. When $k = 0$ (flat spatial geometry), $X_1$, $X_2$ and $X_3$ correspond to the expression, in spherical coordinates, of the usual generators of spatial rotation around respectivelly the $x$-, $y$- and $z$-axis of an orthonormal Cartesian frame, while $X_4$, $X_5$ and $X_6$ correspond to the usual generators of spatial translation along respectivelly the $x$-, $y$- and $z$-axis.

\vspace{10pt}
According to \eqref{eq:Liesym}, a given tensor field $\tensi{T}$ will be cosmologically symmetric provided it satisfies
\begin{equation}
\label{eq:cosmosym}
\Lie_{X_i}\tensi{T} = \tensi{0},\ \forall i= 1,\cdots, 6.
\end{equation}

\section{Technical identities}\label{sec:identity}

In this section, we report for completeness some important relations used in the computations of this paper.

\subsection{Rescaling of the metric}

To perform the variational completion/bootstrapping of our equations, we must know the behavior of the different terms in the equations under the homothety
\begin{equation}
\label{eq:homo}
\chi_u : \left(g_{\mu\nu}, g_{\mu\nu,\rho}, g_{\mu\nu,\rho\sigma}\right) \mapsto \left(u g_{\mu\nu}, u g_{\mu\nu,\rho}, u g_{\mu\nu,\rho\sigma}\right).
\end{equation}

In this respect, one immediately obtains:
\begin{align}
\label{eq:rescale}
\det(g) & \overset{\chi_u}{\longrightarrow} u^{4}\ \det(g)\ , \\
g^{\mu\nu} & \overset{\chi_u}{\longrightarrow} u^{-1}\ g^{\mu\nu}\ , \\
\lc{\Gamma}^\rho_{\mu\nu} & \overset{\chi_u}{\longrightarrow} \lc{\Gamma}^\rho_{\mu\nu}\ , \\
\lc{R}\ud{}{\rho}{\sigma\mu\nu} & \overset{\chi_u}{\longrightarrow} \lc{R}\ud{}{\rho}{\sigma\mu\nu}\ , \\
\lc{R}_{\mu\nu} & \overset{\chi_u}{\longrightarrow} \lc{R}_{\mu\nu}\ , \\
\lc{R} & \overset{\chi_u}{\longrightarrow} u^{-1}\ \lc{R}\ , \\
\lc{G}_{\mu\nu} & \overset{\chi_u}{\longrightarrow} \lc{G}_{\mu\nu}\ , \\
\lc{G}^{\mu\nu} & \overset{\chi_u}{\longrightarrow} u^{-2}\ \lc{G}^{\mu\nu}.
\end{align}

\subsection{Calculus of variations}

To compute the field equations associated to our Lagrangian, we need the following relations.

\subsubsection{Variations with respect to the metric $\metric$}

When computing variation with respect to the metric, one has the usual relations, see \emph{e.g.} \cite{weinberg1972gravitation},
\begin{align}
\delta g_{\mu\nu} = - g_{\mu\alpha} g_{\nu\beta} \delta g^{\alpha\beta}, &\ \ \delta g^{\mu\nu} = - g^{\mu\alpha} g^{\nu\beta} \delta g_{\alpha\beta}, \label{eq:varg}\\
\delta_g \sqrt{-\lvert g\rvert} & = \frac{1}{2} \sqrt{-\lvert g\rvert}\ g^{\mu\nu}\delta g_{\mu\nu}, \label{eq:vardetg}\\
\delta_g\lc{\Gamma}^{\rho}_{\mu\nu} & = \frac{1}{2} g^{\rho\alpha} \left(\lc{\nabla}_\mu \delta g_{\nu\alpha} - \lc{\nabla}_\alpha \delta g_{\mu\nu} + \lc{\nabla}_\nu \delta g_{\alpha\mu} \right),\label{eq:varGamma}\\
\delta_g \lc{R}\ud{}{\rho}{\sigma\mu\nu} & = \lc{\nabla}_{\mu} \delta_g\lc{\Gamma}^{\rho}_{\sigma\nu} - \lc{\nabla}_{\nu} \delta_g\lc{\Gamma}^{\rho}_{\sigma\mu},\label{eq:varR}\\
\delta_g\left(\lc{\nabla}_\alpha \ud{L}{\rho}{\sigma\beta}\right) & = \delta_g\lc{\Gamma}^{\rho}_{\mu\alpha} \ud{L}{\mu}{\sigma\beta} - \delta_g\lc{\Gamma}^{\mu}_{\sigma\alpha} \ud{L}{\rho}{\mu\beta} - \delta_g\lc{\Gamma}^{\mu}_{\beta\alpha} \ud{L}{\rho}{\sigma\mu}.\label{eq:gvarL}
\end{align}

\subsubsection{Variations with respect to the distortion $\tens{L}$}

When computing variation with respect to the distortion, we get:
\begin{align}
\delta_L\left(\lc{\nabla}_\alpha \ud{L}{\rho}{\mu\nu}\right) & = \delta_L\left(\partial_\alpha \ud{L}{\rho}{\mu\nu} + \lc{\Gamma}^{\rho}_{\beta\alpha} \ud{L}{\beta}{\mu\nu} - \lc{\Gamma}^{\beta}_{\mu\alpha} \ud{L}{\rho}{\beta\nu} - \lc{\Gamma}^{\beta}_{\nu\alpha} \ud{L}{\rho}{\mu\beta}\right)\\
& = \partial_\alpha \left(\delta\ud{L}{\rho}{\mu\nu}\right) + \lc{\Gamma}^{\rho}_{\beta\alpha} \delta\ud{L}{\beta}{\mu\nu} - \lc{\Gamma}^{\beta}_{\mu\alpha} \delta\ud{L}{\rho}{\beta\nu} - \lc{\Gamma}^{\beta}_{\nu\alpha} \delta\ud{L}{\rho}{\mu\beta}\\
& = \lc{\nabla}_\alpha \left(\delta\ud{L}{\rho}{\mu\nu}\right).\label{eq:varL}
\end{align}

\end{document}